# Relationship between cellular response and behavioral variability in bacterial chemotaxis

Thierry Emonet* & Philippe Cluzel

*The Department of Physics, The Institute for Biophysical Dynamics, and The James Franck Institute, University of Chicago, Chicago, IL, USA*

*current address: MCDB, Yale University, P.O. Box 208103, New Haven CT, 06520

Classification: Biological Sciences / Biophysics
Corresponding authors: Thierry Emonet (thierry.emonet@yale.edu) and Philippe Cluzel (cluzel@uchicago.edu)

**Bacterial chemotaxis in *Escherichia coli* is a canonical system for the study of signal transduction. A remarkable feature of this system is the coexistence of precise adaptation in population with large fluctuating cellular behavior in single cells (Korobkova et al. 2004, Nature, 428, 574). Using a stochastic model, we found that the large behavioral variability experimentally observed in non-stimulated cells is a direct consequence of the architecture of this adaptive system. Reversible covalent modification cycles, in which methylation and demethylation reactions antagonistically regulate the activity of receptor-kinase complexes, operate outside the region of first-order kinetics. As a result, the receptor-kinase that governs cellular behavior exhibits a sigmoidal activation curve. This curve simultaneously amplifies the inherent stochastic fluctuations in the system and lengthens the relaxation time in response to stimulus. Because stochastic fluctuations cause large behavioral variability and the relaxation time governs the average duration of runs in response to small stimuli, cells with the greatest fluctuating behavior also display the largest chemotactic response. Finally, Large-scale simulations of digital bacteria suggest that the chemotaxis network is tuned to simultaneously optimize the random spread of cells in absence of nutrients and the cellular response to gradients of attractant.**

Bacterial chemotaxis is a system that controls the locomotion of cells and has arisen as a model system for signal transduction. Although the chemotaxis network is composed of just a few molecular species, it can perform complex cellular functions such as adaptation in response to environmental changes. *E. coli* bacteria swim using flagella activated by rotary motors. When motors spin counterclockwise (CCW), the flagella form a corkscrew bundle whose rotation propels the cell in a smooth trajectory called a run. Clockwise (CW) spinning of motors favors the disruption of the bundle and causes the cell to tumble. The trajectory of a swimming cell resembles that of a random walk, which consists of a succession of runs and tumbles. When a bacterium is exposed to a sudden increase of chemical attractant, it lengthens the time interval between two consecutive tumbles. This modulation of tumbling rate allows bacteria to bias their random walk toward a source of attractants and to perform chemotaxis.

A combination of experiments and models demonstrated that at the population level *E. coli* bacteria display exact adaptation: after an initial response to a chemical stimulus, the average tumbling rate of bacteria returns precisely to its steady pre-stimulus level (1-3). Remarkably, these studies showed that the property of exact adaptation is robust to variations of biochemical parameters such as the concentration and reaction rates of the chemotaxis proteins (2, 3). In contrast with population measurements, recent single-cell experiments have revealed that the switching behavior of a single motor, from cells adapted to a homogeneous environment, exhibits large temporal fluctuations (4, 5). It was found that the characteristic time scale of these fluctuations was so large that a steady tumbling rate could be defined only when taking time averages longer than several hundreds seconds. The amplitude of these fluctuations, hereafter called noise, exceeded that expected from Poisson statistics. Very long time series of switching events from individual flagellar motors exhibited distributions of run (CCW) intervals with long tails and that are not exponentially distributed as it was previously believed (4, 5). The role of molecular noise (i.e. stochastic fluctuations) as a source of phenotypic variability has recently been reported in a number of biological systems as diverse as gene expression and signal transduction in prokaryotes and eukaryotes (6-8). It is conceivable, that biological systems that are sensitive to intracellular spontaneous noise are also sensitive to small extra-cellular perturbations such as a sudden change of environmental conditions. Here, we develop a stochastic model of adaptation in bacterial chemotaxis to analyze how the noise from non-stimulated cells could be related to the cellular response to a chemical stimulus. This model shows how the adaptation mechanism that drives the chemotactic response may also amplify the spontaneous fluctuations taking place in the signaling pathway.

**Results**

**Kinetics of the adaptation module**

The adaptation system in bacterial chemotaxis is governed by a well-defined module of specific interacting proteins. This adaptation module consists of methylation/demethylation cycles in series that control the activity of a histidine kinase CheA (Figure 1). Two antagonistic enzymes, CheR and CheB-P, respectively add and remove methyl groups at multiple receptor residues of the receptor-kinase complex. We use a standard two-state model that yields exact adaptation. Each receptor complex can be either active or inactive (2, 9-12) (Figure 1B) and Michaelis-Menten kinetics governs the methylation-demethylation steps of the receptor complexes. We will distinguish the concentration of free receptor complexes from the concentrations of the intermediate compounds, the receptor-CheR and receptor-CheB-P. Under these conditions, the temporal evolution of the average methylation level, $M$, of the free receptors throughout the methylation / demethylation cycles (12) obeys Eq. (1) (Materials and Methods):

$$\frac{dM}{dt} = \underbrace{\frac{k_r \varepsilon_r}{K_r + A}}_{r} A - \underbrace{\frac{k_b \varepsilon_{bp}}{K_b + A^*}}_{b} A^* \qquad (1)$$

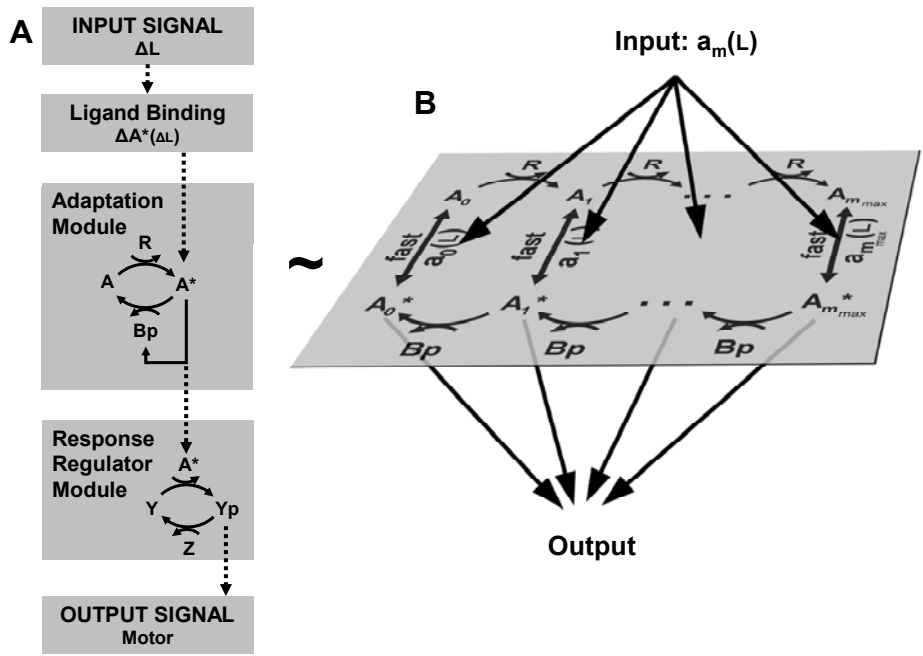

**Figure 1:** Modular representation of the chemotaxis system. (**A**) Transmembrane receptors bind the ligand ($L$) and control the activity of histidine kinases CheA ($A$). The kinase CheA phosphorylates the response regulator CheY ($Y$) into the active form CheY-P ($Y_p$). CheY-P diffuses throughout the cell and interacts with the flagellar motors to induce clockwise rotation (tumble). The phosphatase CheZ ($Z$) dephosphorylates CheY-P (34). A sudden increase of ligands $\Delta L$ causes the kinase activity to decrease by $\Delta A^*$. The chemotaxis system is equipped with an adaptation module in which two antagonistic enzymes regulate the activity of the kinase-receptor complexes. The methyltransferase CheR ($R$) catalyzes the autophosphorylation of CheA by methylating the receptors. The active kinase $A^*$ phosphorylates the methylesterase CheB in CheB-P ($B_p$). CheB-P removes methyl groups from active receptor complexes, which catalyzes kinase deactivation. (**B**) The adaptation module consists of a series of slow (de)methylation reactions that modulate the activity of the receptor complexes. We use a two-state model where the probability $a_m$ of a receptor complex to be in active conformation depends on the occupancy of its ligand binding sites and on the level of methylation of the receptors that ranges within $m = 0,\ldots,m_{max}$ (2, 11, 12, 27). $m_{max}$ is the total number of methylation sites. We assume that CheR only methylates inactive complexes (35), whereas CheB-P only demethylates active complexes (36) (details on the model and alternative hypotheses in Supporting information).

Here $A^*$ and $A$ are the concentrations of free active and inactive receptor complexes, and $r$ and $b$ are the rates of methylation and demethylation of inactive and active receptors complexes. The parameters $\varepsilon_r$ and $\varepsilon_{bp}$ are the concentrations of CheR and CheB-P, $K_r$ and $K_b$ are effective Michaelis-Menten constants for the methylation-demethylation of a receptor complex, and $k_r$, $k_b$ are the corresponding catalytic rates. The total concentration of receptor complexes is constant.

$$\underbrace{A\left(1+\frac{r}{k_r}\right)}_{A_{tot}} + \underbrace{A^*\left(1+\frac{b}{k_b}\right)}_{A^*_{tot}} = 1 \qquad (2)$$

The first (second) term in equation (2) represents the total concentration of inactive (active) receptor complexes including those bound to CheR and CheB. The right hand side of equation (2) is equal to 1, because we normalize all the concentrations ($A$, $\varepsilon_r$, $K_r$, etc.) with the concentration, $N$, of receptor complexes in the system. At steady state $dM/dt = 0$ and equations (1) and (2) yield the steady state concentrations of active and inactive receptor complexes, $\bar{A}^*_{tot}$ and $\bar{A}_{tot}$. The concentration of ligand in the external medium does not appear explicitly in Eqs. (1) and (2) and therefore $\bar{A}_{tot}$ is independent from the steady state concentration of ligand in the external medium (Materials and Methods). Thus, within this kinetic approach, the adaptation module exhibits perfect adaptation (2, 11).

The equation that governs the steady state of the kinase activity in the chemotaxis system was calculated from Eqs. (1) and (2). Importantly, we find that this equation is identical to the steady state equation obtained by Goldbeter and Koshland for covalent modification of a common substrate by two antagonistic enzymes (compare Fig. 3 in ref. (13) with Eq. (S15)). Thus, the adaptation module (Figure 1) in chemotaxis may share some of the properties of a modification cycle. As in ref. (13), the normalized constants, $K_r$ and $K_b$ control the steepness of the activation curve of the kinase as a function of the modifying enzymes. When $K_r$ and/or $K_b$ are smaller than one, the activation curve is steeper than a hyperbolic function (Hill coefficient >1). The ratio $\alpha = k_r \varepsilon_r / k_b \varepsilon_{bp}$ of the maximal enzymatic velocities determines the steady state activity of the kinase (13). Because the ratio $\alpha$ is proportional to the ratio of [CheR] and [CheB-P], it can be easily adjusted by changing the relative concentration of these two proteins.

Using the parameter values from recent experimental data (Table S1), we plotted the steady state activity of the kinase, $\bar{A}^*_{tot}$, as a function of [CheR] for a fixed wild type level of [CheB] (Figure 2A). In the absence of the CheB-P feedback loop, the kinase activation curve exhibits an effective Hill coefficient of 3.5. In presence of the CheB-P feedback loop, the effective Hill coefficient is 2.5 (Figures 2B). This feedback loop also helps maintaining the kinase activity within the narrow functioning range of the rotary motor (14). The activation curve of the kinase activity as a function of 1/[CheB], and at fixed wild type level of [CheR], exhibits a Hill coefficient larger than one as well (data not shown).We also calculated the kinase activation curve for other models, such as the stochastic numerical model of chemotaxis developed independently by Bray and col.(10) and for a model where CheR methylates the receptors in both active and inactive conformations. In all cases we find that the Hill coefficients are larger than one (Supporting Information). These results suggest that the adaptation module in bacterial chemotaxis is more sensitive to small variations of the catalytic activity of CheR and CheB-P than an equivalent system with a hyperbolic activation curve. This sensitivity also depends on the ratio $\alpha$ of the methylation and demethylation velocities. Outside the transition region of the sigmoid (Figure 2B), the kinase would be either fully active or fully inactive. Thus, we expect the wild type value of $\alpha$ to be adjusted about the transition region of the sigmoid where the activity of the kinase is allowed to vary (like Bray and col. (10).)

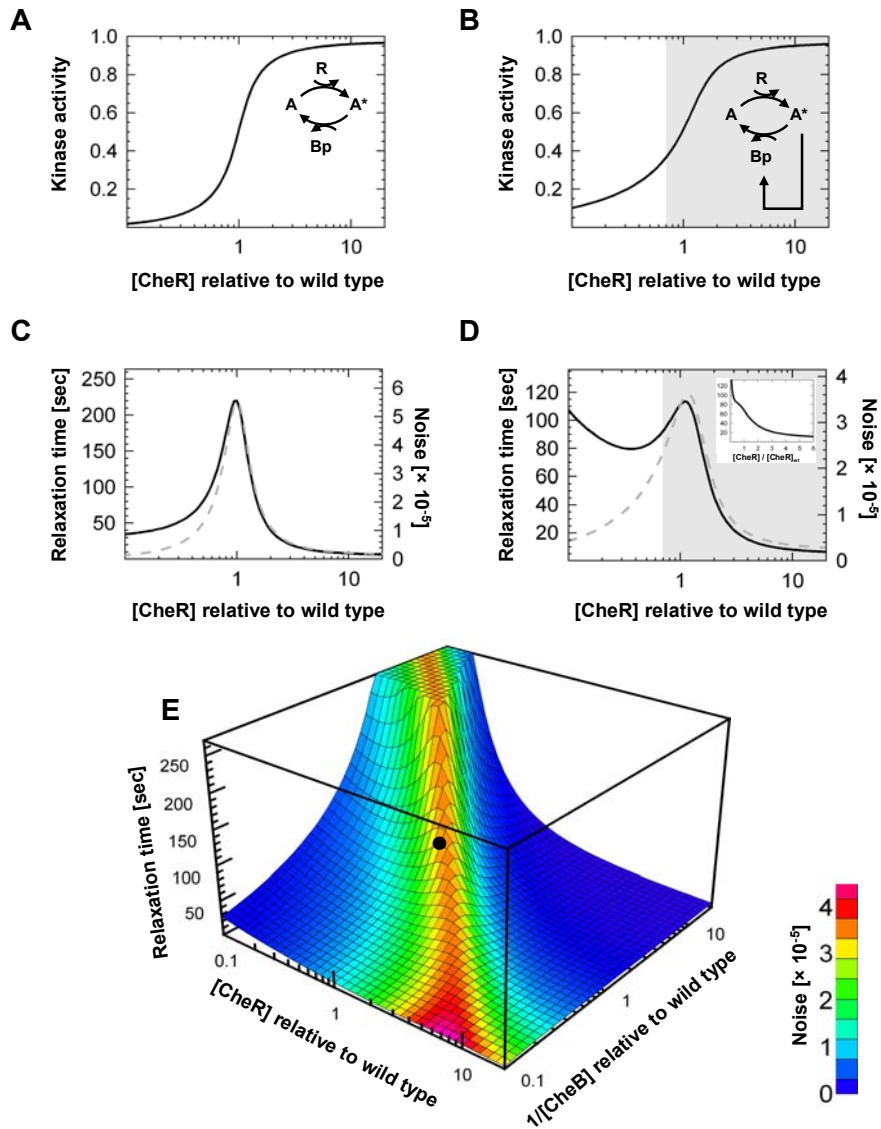

**Figure 2:** Sensitivity of the adaptation module without (A, C) and with (B, D, E) the CheB-P feedback loop on CheA. (**A**) (black) Total kinase activity $A^*_{tot}$ as a function of [CheR] for a fixed wild type level of [CheB] (model parameters in Table S1). Hill coefficient, $H \sim 3.5$. (**C**) Relaxation time $\tau_a$ (black) and variance $\sigma^2_a$ (grey) of the noise associated with the total kinase activity. (**B**, **D**) Same as (A, C) with the CheB-P feedback loop on CheA. $H \sim 2.5$. (Grey-shaded area) Functioning range of the motor ($1.5 <$ [CheY-P] $< 4.5$ μM (37)). (Inset) Relaxation time averaged from a population of cells ($10^4$) with cell-to-cell variability in the levels of CheR and CheB (Supporting Information). (**E**) Relaxation time (surface) and variance of the noise (color) associated with the total kinase activity as a function of [CheR] and 1/[CheB]. (●) wild-type cell (Table S1).

**Stochastic model of the adaptation module**

Recent experiments showed that individual *E. coli* bacteria adapted to a homogeneous environment exhibited large temporal variations in their behavior (5). It was hypothesized that this behavioral variability is due to the fluctuations of the kinase activity governed by the slow methylation-demetylation process in the adaptation module. Three complementary experiments in which cells did not display large fluctuations at long time scales support this hypothesis: (i) when [CheY-P] was not regulated by the chemotaxis network, but substituted by the active mutant CheYD13K stably expressed from an inducible plasmid; (ii) In mutant cells whose receptors have a fixed methylation level; (iii) behavioral variability was found tunable in Δ*cheR* mutant cells complemented with various levels of [CheR]. Behavioral variability decreased from its maximal to minimal value when [CheR] was varied from one to fourfold wild type level.

This section aims to interpret these experiments using a stochastic model of the adaptation module. In particular, we discuss steepness of the activation curve (Hill coefficient >1) of the kinase (Fig. 2B) could amplify the spontaneous stochastic fluctuations associated with the methylation and demethylation reactions. We wish to show that the amplification of fluctuations in kinase activity is large enough to produce the large behavioral variability observed experimentally in non-stimulated cells (5). Commonly, chemotaxis assays are performed in media that do not support growth but provide the essential elements in excess to produce energy. Consequently, cells can be observed for several hours without noticeable metabolic changes (15). We then consider that bacteria that are adapted to a homogenous environment are at the equilibrium. Under this condition, thermal fluctuations cause spontaneous fluctuations in the rate constants of the chemical reactions taking place in the chemotaxis network (16, 17).

We use a linear noise approximation (16, 18-20) to derive the spontaneous noise in the kinase activity (and in [CheY-P]) around the steady state (Figs. 2A-B). We assess the validity of this linear noise approximation by comparing its results from that of full stochastic numerical simulations (Supporting Information). This approach allows us to characterize the relaxation time, $\tau_a$, and the power spectrum of the spontaneous fluctuations in the output signal of the chemotaxis system. The relaxation time provides the time scale over which the steady state of a non-stimulated cell is defined. At time scales shorter than $\tau_a$, the spontaneous fluctuations of the system are strongly correlated and the behavior is unsteady.

For sake of simplicity, we first ignore the phosphorylation step of CheB. The spontaneous fluctuations in kinase activity, $\delta A^*$, about the steady state value $\overline{A}^*$, include 'fast' fluctuations of the receptors activity due to the binding and unbinding of ligand, and also slow fluctuations, $\delta M / m_{max}$, associated with the methylation-demetylation reactions ($m_{max}$ is the maximal number of methylation sites). At long time scales, only the slow fluctuations are relevant and $\delta A^* \cong \delta M / m_{max}$ (Materials and Methods). Inserting the latter relation in equation (1), and using the linear noise approximation yields the noise, $\delta A^*_{tot}$, around the steady state, $\overline{A}^*_{tot}$, associated with the total kinase activity in non-stimulated cells.

$$\frac{d}{dt}\delta A_{\text{tot}}^{*} = -\frac{1}{\tau_a}\delta A_{\text{tot}}^{*} + \sqrt{D_a}\,\delta\eta_a \tag{3}$$

The last term of this Langevin Eq. (3) represents the spontaneous fluctuations associated with (de)methylation reactions of the receptors. $\delta\eta_a$ is a source of stochastic white noise. The Langevin equation (3) is analogous to the equation $\gamma\dot{x} = -\kappa x + f(t)$ that describes the fluctuations of a mass-spring system in a viscous fluid, with spring constant $\kappa = k_B T N m_{\max}/(\partial \overline{A}^*/\partial \ln \varepsilon_r)$, damping constant $\gamma = k_B T N m_{\max}^2/(\overline{b}\,\overline{A}^*)$ and fluctuating force $f(t)$. Fluctuations and dissipation are related through the fluctuation-dissipation theorem $\langle f(t)f(t')\rangle = 2k_B T \gamma \delta(t-t')$, where $T$ is the temperature and $k_B$ the Boltzmann constant (16). As a consequence, we find that the strength of the spontaneous fluctuations associated with (de)methylation reactions is

$$D_a \cong \frac{\overline{r}\,\overline{A} + \overline{b}\,\overline{A}^*}{N m_{\max}^2} = \frac{2\overline{b}\,\overline{A}^*}{N m_{\max}^2} \tag{4}$$

and the variance of the spontaneous fluctuations in kinase activity $\delta A_{\text{tot}}^{*}$ is $\sigma_a^2 = k_B T/\kappa = \tau_a D_a/2$. Interestingly, $D_a$ is independent from the slope of the kinase activation curve ($\overline{b}\,\overline{A}^*$ is approximately $k_b \overline{\varepsilon}_{bp}/(1+2K_b) \cong k_b \overline{\varepsilon}_{bp}$ in the middle of the transition). By contrast, the relaxation time, $\tau_a$, is proportional to the slope (i.e. the Hill coefficient) of the kinase activation curve:

$$\tau_a(\varepsilon_r) \cong \frac{m_{\max}}{\overline{b}\,\overline{A}^*}\left(\frac{\partial \overline{A}^*}{\partial \ln \varepsilon_r}\right)_{\varepsilon_b} \tag{5}$$

Thus, increasing the Hill coefficient of the kinase activation curve is equivalent to decreasing the spring constant $\kappa$ in the mass-spring system without changing the damping constant $\gamma$. The higher the Hill coefficient, the larger the fluctuations of the kinase activity and the longer the relaxation time. Importantly, we find that $\tau_a$ and $1/\sigma_a^2$ are proportional to the relaxation time and the inverse of the noise for only one covalent modification cycle (e.g. the Goldbeter & Koshland system (13)). The proportionality factor is $m_{\max}$.

A practical way to study the behavioral variability of an *E. coli* bacterium is to plot the power spectrum associated with the fluctuations in the output of the full chemotaxis system when the cell is adapted to its environment. Performing the Fourier transform of Eq. (3), we obtain the power spectrum of the spontaneous fluctuations in the kinase activity. Similarly, we also calculate the power spectrum of the fluctuations of [CheY-P] in the full pathway that includes the CheB-P feedback loop (Supporting Information). In Fig. 3, we plot the power spectra of the fluctuations of [CheY-P] for various values of [CheR] and the Michaelis-Menten constants $K_r$ and $K_b$. Using recently published biochemical parameters (21), we find that this model best reproduces single cells experiments (power spectra) (5) when the two following conditions are simultaneously fulfilled: (i) $K_r$ and $K_b$ are about $10^{-1}$ and (ii) the methylation and demethylation catalytic rates $k_r$ and $k_b$ are adjusted such that wild type cells, like in experiments, exhibit the noisiest behavior (Table S1). The values of these parameters are also in agreement with biochemical data and parameters used in earlier models (Supporting Information). By contrast, when $K_r, K_b \approx 1$, we find that the behavioral variability is not sensitive to variations of [CheR] as it was found in experiments (5). Similarly, when $K_r, K_b \approx 10^{-2}$, the difference in behavioral variability between wild type and mutants cells that express four fold wild type level of [CheR] is too large in comparison with experiments (5).

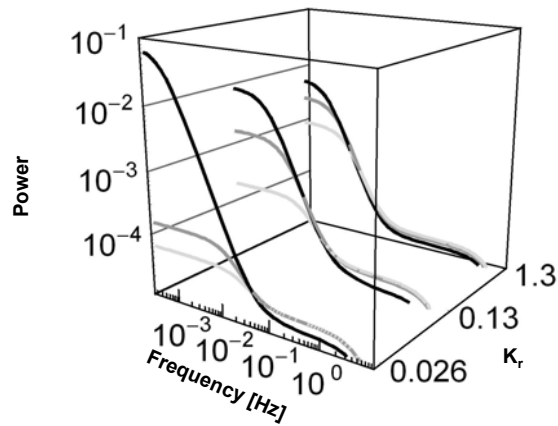

**Figure 3:** Power spectra of the fluctuations of output signal (CheY-P) from non-stimulated cells. One (black), two (grey) and four fold (light grey) wild-type levels of CheR for a fixed wild type level of [CheB]. For $K_r = 0.132$ and $K_b = 0.176$, the spectra are in agreement with experiments in (5). For $K_r$ and $K_b$ five times smaller or ten times larger, the differences between the power spectra of wild type and mutants are too large or too small in comparison with those in ref. (5)

The power spectra in Fig. 3 exhibit two characteristic (knee) frequencies. At low frequency, the slow fluctuations of the kinase activity that are governed by the adaptation module dominate. The position of the knee at long time scale reflects the magnitude of $\sigma_a^2$ and $\tau_a$, which depend on [CheR] and [CheB-P]. This knee frequency can also be interpreted as a low cutoff frequency of the chemotaxis system (Material and Methods and ref. (15)). At short time scale, the second knee frequency corresponds to the spontaneous fluctuations occurring within the phosphorylation cascade and its position depends only on the timescale associated with the phosphorylation of CheY (Figure 1A). Next, we plotted the noise in kinase activity, $\sigma_a^2$, and the relaxation time, $\tau_a$, as a function of [CheR], in absence or presence of the CheB-P feedback loop (Fig. 2C, 2D). As expected for a covalent modification cycle with a Hill coefficient larger than one (13, 22-24), $\sigma_a^2$ and $\tau_a$ peak within the transition region of the kinase activation curve (Fig. 2C). This effect is similar to the one observed in (22, 24). In presence of the CheB-P feedback loop, the overall profiles of the variance and relaxation time as a function of [CheR] are conserved, with the exception of $\tau_a$ that decreases at very low level of CheR (Fig. 2D). We also find that the CheB-P feedback loop slightly reduces the noise and the relaxation time. We obtain similar results when we plot $\sigma_a^2$ and $\tau_a$ as a function of 1/[CheB]. In Figure 2E, we plot $\sigma_a^2$ and $\tau_a$ as a function of both [CheR] and 1/[CheB] for the whole system including the CheB-P feedback loop. The presence of the crest over the hyperbolic surface is the signature of the deviation from first-order kinetics. As expected, large fluctuations in kinase activity coincides with long relaxation times.

Figures 2 and 3 suggest that in wild type cells, the ratio $\alpha$ of the methylation and demethylation velocities must be tuned within the transition region of the activation curve of the kinase (Fig. 2B) in order to exhibit fluctuations (Fig. 3) similar to those observed in (5). Furthermore, in order to have these fluctuations sensitive to small variations in [CheR], as observed in experiments, the activation curve of the kinase activity (Fig. 2B) must exhibit a Hill coefficient larger than one. These predictions are in close agreement with experiments (power spectra) in *ΔcheR* mutant cells complemented with various levels of CheR. Finally, this model also predict the experimental observation on single cells, which show that the behavioral variability is maximal when CheR is expressed at wild type level, and decreases to the variability expected from Poisson statistics when [CheR] is expressed at fourfold wild type level (5).

**Relationship between behavioral variability in non-stimulated cells and the timing of the chemotactic response to small stimuli**

The critical hypothesis in this section is that the behavioral variability observed in *non-stimulated* single cells is fundamentally related to the timing of the response in *stimulated* cells. Because stimuli encountered by the bacterium in its natural habitat may be small, we calculate the response to stimulus using a linear perturbation analysis of the kinetic system. Although we chose to keep our study independent from the actual chemical stimulus present in the environment, the chemotactic response in the real system also depends on the initial amplification of the input stimulus mediated by complex allosteric mechanisms taking place at the level of the receptors (25-29). In this model, we consider that a small external perturbation, such as a sudden exposure to attractant, causes an 'instantaneous' change of the receptors activity, $\Delta A^*_{\text{input}}$. During the subsequent adaptation, the changes $\Delta M$ in methylation levels of the receptors govern the changes in kinase activity. We linearize Eq. (1) and eliminate $\Delta M$ (Material and Methods) to obtain the input-output relationship of the adaptation module solely in terms of changes in total kinase activity $\Delta A^*_{\text{input}}$:

$$\frac{d}{dt}\Delta A^*_{\text{tot}} = -\frac{1}{\tau_a}\Delta A^*_{\text{tot}} + \frac{d}{dt}\Delta A^*_{\text{input}} \qquad (6)$$

Within this linear approximation, the sensory chemotaxis pathway relaxes with the same time scale $\tau_a$ in response to either an external stimulus or internal spontaneous noise (Eqs. 3 and 6). Thus, one can infer the sensitivity of the chemotactic response $\Delta A^*_{\text{tot}}(t)$ to a *small* external perturbation $\Delta A^*_{\text{input}}$ by characterizing the time correlation of the noise, $\delta A^*_{\text{tot}}$, in non-stimulated cells. In this picture, a large $\tau_a$ will cause on average long runs following a *small* step stimuli $\Delta A^*_{\text{input}}$. Consequently, we interpret the relaxation time, $\tau_a$, as a relative measure of the sensitivity of the chemotaxis system in response to a small external perturbation $\Delta A^*_{\text{input}}$. Thus, we expect the chemotactic response to peak with the relaxation time $\tau_a$ at about wild-type level of [CheR] and 1/[CheB] (Figs. 2D-2E). For small or large values of [CheR] and/or 1/[CheB], it is conceivable that [CheY-P] becomes too large to fall within the narrow functioning range of the motor. Under this extreme condition, the motor doesn't switch and the system is not chemotatic. In population measurements (3), it is difficult to observe the peaking of the relaxation time because of the inherent cell-to-cell variability of *cheR* and *cheB* expression levels (Figure 2D, inset and Supporting Information Section 4). But this prediction should be testable by the means of single cell experiments.

In order to highlight the significance of the relaxation time for chemotaxis, we hypothesize that the average duration of runs varies like the time $\tau_a$ when cells respond to small input stimuli (Fig. 4A). We anticipate that the chemotactic drift associated with cells swimming upward gradient of nutrients will be larger for wild-type cells with a larger $\tau_a$ than that of CheR mutants (Fig. 4B). We tested this hypothesis by performing large scale computer simulations with the agent-based simulator AgentCell (30) (http://www.agentcell.org). We find that the chemotaxis response decreases with small variations in [CheR] relative to wild-type level (Figure 4C). Small changes in the concentration of CheB from the wild-type level produced similar results (Fig.S7). To check that this result is not due to [CheY-P] lying outside of the functioning range of the motor, we adjusted the narrow functioning range of the motor so that the CW bias would remain the same in all populations expressing various level of [CheR] (CW=0.23).

Since wild-type cells display the largest behavioral variability, wild-type cells spread further than mutant cells in the direction perpendicular to the gradient of nutrient (figure 4D). It is then conceivable that the non-linearity (i. e. Hill coefficient >1) of the adaptation module (5) confers two advantages: First, it enhances the chemotaxis drift by producing long runs along a given gradient of attractant. Second, it produces a large behavioral variability that allows a population of bacteria to explore a wider area. These two predictions should be testable experimentally. In real bacterial populations, cell-to-cell variations of [CheR] and [CheB] arise naturally. Given the fine-tuning of the chemotactic response on [CheR] and [CheB-P], we anticipate the existence of cellular mechanisms to maintain the value of the ratio of the methylation-demethylation velocities within a functioning range. For example, *cheR* and *cheB* genes are adjacent on the multi-cistronic *meche* operon, a strategy known to reduce independent variations between the expression of the two genes (Figure S4). Another mechanism that helps maintaining the [CheR]/[CheB-P] ratio is the existence of the negative feedback loop on CheB as illustrated in (14).

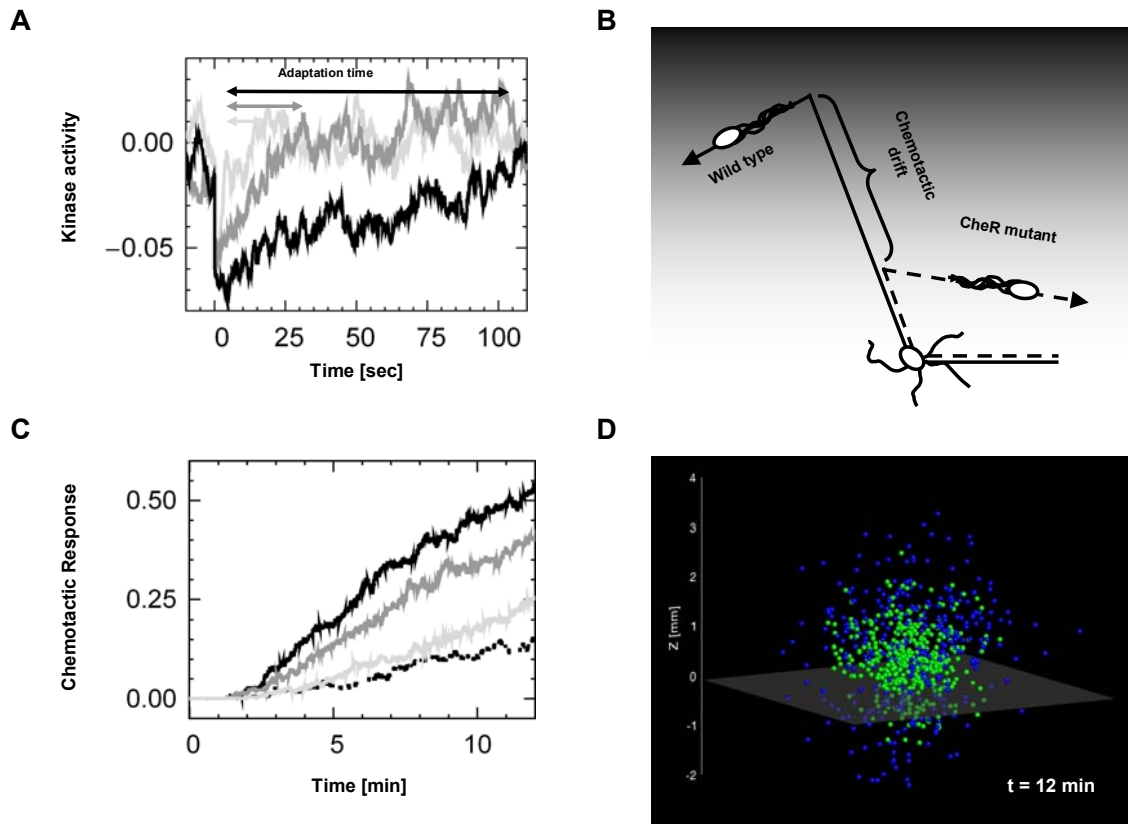

**Figure 4:** Relationship between relaxation time and chemotactic drift. (**A**) Temporal evolution of the kinase activity relative to steady state upon sudden deactivation of active receptor complexes for one (black), two (grey), and fourfold (light grey) wild-type level of [CheR] and fixed wild-type level of [CheB]. We normalized the kinase activity with that of wild type cells. Increasing [CheR] causes a reduction of the relaxation time $\tau_a$. In all cases, the initial perturbation is ~4.4% of the wild type steady state kinase activity (~100 receptor complexes). Stochastic simulations of a single covalent modification cycle (Fig. 1A) using Stochastirator (http://opnsrcbio.molsci.org/); parameters in Table S1. (**B**) Role of the relaxation time in chemotaxis. Cells with longer relaxation time swim farther along the gradient of attractant (gray shade). (**C**) Effect of variations of [CheR] on the chemotactic response of a bacterial population of 400 cells. Digital swimming bacteria are exposed to a constant gradient of aspartate ($dL/dz = 10^{-8}$ M/μm, $L(z=0) = 1$ μM). Percentage of cells in the gradient that are above $z = 1$ mm as a function of time: one (black), two (grey), four (light grey) wild-type [CheR] level. (Dashed line) wild type response without gradient. The CW bias for mutant and wild-type cells is 0.23. The initial position of the bacteria is $z = 0$ mm. (**D**) Position of the cells from (C) with one (Green) and fourfold (Blue) wild type level of CheR after 12 minutes. Below the grey transparent plane ($z = -0.1$ mm) there is no nutrient and bacteria perform an unbiased random walk. Above the plane the random walk is biased upward the gradient of aspartate.

This stochastic approach to adaptation in bacterial chemotaxis reconciles the presence of large behavioral variability observed at the single cell level with the chemotactic response of a cellular population. From the design analysis of the chemotaxis system, we showed that the non-linearity of the adaptation mechanism simultaneously amplifies both the noise and the relaxation time in the chemotaxis system. This relationship between noise and relaxation time allows us to predict that the noisiest cells should exhibit the largest chemotactic response. We illustrated this relationship with large-scale simulations of digital bacteria swimming upward a gradient of attractant. The ability to infer a particular cellular response from behavioral variability in non-stimulated cells may be extended to theoretical and experimental studies of other signaling cascades. This approach highlights a key property of the adaptive system in chemotaxis, which consists of a relationship between cellular response and behavioral variability.

**Materials and Methods**

**Details of the analytical model.** We combined the two-state model of receptors in (9) with the exact adaptation mechanism in (2). The phosphorylation cascade is similar to (26). The resulting kinetic model is essentially the same as in (10-12, 14). Derivation of the full kinetic and stochastic models is available in Supporting information.

**Kinetic system.** Ligand binding and conformational changes are much faster than the (de)methylations reactions. Therefore, we describe the activity of receptor complexes using equilibrium probabilities $a_m(L)$ (Fig. 2B) that are functions of the concentration of ligand in the external medium. The subscript $m = 0, \ldots, m_{\max}$ indicates the methylation level of the receptor complex. The methylation-demethylation steps and the time evolution of the concentrations $X_m$ of free (not bound to enzyme) receptor complexes with $m$ methyl groups (Fig. 2B) are governed by

$$\frac{d}{dt} X_m = r\left[(1-a_{m-1})X_{m-1} - (1-a_m)X_m\right] + b\left[a_{m+1} X_{m+1} - a_m X_m\right] \quad (7)$$

Multiplying Eq. (7) with $m$ and summing over all $m$'s gives Eq. (1), where $A^* = \sum_m a_m X_m$ and $A = \sum_m (1-a_m) X_m$.

**Perturbation analysis of the adaptation module**. The changes of kinase activity $\Delta A^*$ are:

$$\Delta A^*(t) = \underbrace{\sum_{m=0}^{m_{\max}} \Delta a_m(t)\, \overline{X}_m}_{\Delta A^*_{\text{input}}} + \underbrace{\sum_{m=0}^{m_{\max}} \overline{a}_m \Delta X_m(t)}_{\text{Adaptation}} \quad (8)$$

To keep our analysis independent from various models of receptors, we use directly $\Delta A^*_{\text{input}} = \sum_m \Delta a_m\, \overline{X}_m$ as the input of the adaptation module without detailing the relationship of $\Delta a_m$ to changes in ligand concentration (Fig. 1B). In previous models of bacterial chemotaxis (10, 12, 14), the probability of activation of receptors complexes at steady state increases approximately linearly with the number of methyl groups when the external concentration of ligand is small (Fig. S6). In our linear perturbation analysis we take $\bar{a}_m \approx m/m_{max}$. The second term in Equ. (8) therefore becomes $\Delta M / m_{max}$. This linear approximation simplifies the analytical treatment while capturing the basic dependence of the kinase activity on methylation level as established by biochemistry data(14, 26, 31). We validate the approximation by comparing the analytical results with results from stochastic simulations that include non-linear $a_m(L)$ (Supporting Information). When considering the *slow* stochastic fluctuations in non-stimulated cell we have: $\delta A^* = \delta A^*_{input} + \delta M / m_{max} \cong \delta M / m_{max}$.

**Relaxation time.** At steady state, the total derivative of the equilibrium relation $\bar{r}\bar{A} = \bar{b}\bar{A}^*$ yields $\tau_{GK}^{-1} d\bar{A}^* = \bar{r}\bar{A} d\ln\varepsilon_r - \bar{b}\bar{A}^* d\ln\bar{\varepsilon}_{bp}$, where $\tau_{GK}$ is the relaxation time for only one modification cycle (e.g. the Goldbeter & Koshland system(13)). We neglected the small corrections due to the receptor complexes bound to CheR and CheB-P (the full derivation is in Supporting Information). In the system with the CheB-P feedback loop, $\bar{\varepsilon}_{bp}$ is a function of the kinase activity $\bar{A}^*$ and of $\varepsilon_b$. Thus, $d\ln\bar{\varepsilon}_{bp} = \mu_a d\ln\bar{A}^* + \mu_b d\ln\varepsilon_b$ which, together with the previous equation, yields equation (4) with $\tau_a^{-1} = (\tau_{GK}^{-1} + \bar{b}\bar{A}^*\mu_a)/m_{max}$. Here, $\mu_a = (\partial\ln\bar{\varepsilon}_{bp}/\partial\ln\bar{A}^*)_{\varepsilon_r}$ represents the strength of the CheB-P feedback loop and $\mu_b = (\partial\ln\bar{\varepsilon}_{bp}/\partial\ln\varepsilon_b)_{\bar{A}^*}$. Similarly to Eq. (5) we have $\tau_a(\varepsilon_b^{-1}) \cong m_{max}/(\bar{b}\mu_b) \left(\partial\ln\bar{A}^*/\partial\ln\varepsilon_b^{-1}\right)_{\varepsilon_r}$.

**Filtering properties of the pathway.** The equation (6) represents a negative integral feedback system. From the Fourier transform of Eq. (6), we find that the frequency response of the adaptation module behaves like a high pass filter with a cutoff frequency $\tau_a^{-1}$. Similarly, the frequency response of the response regulator module (Figure 1A) behaves like a low pass filter with a cutoff frequency $\tau_y^{-1}$, where $\tau_y$ is the relaxation time of the response regulator module (Supporting Information). Thus, the chemotaxis system resembles a bandpass filter. The value of the low cutoff frequency $\tau_a^{-1}$ predicted by our model for wild type level of [CheR] is in agreement with measurements from tethered cells (15). A recent analysis of the noise filtering properties of the chemotaxis system suggested that lower cutoff frequency $\tau_a^{-1}$ corresponds to longer time integration of the input signal and therefore better removal of high frequency fluctuations (32). This study however did not include the slow spontaneous fluctuations of the kinase activity due to stochastic fluctuations in the methylation-demethylation process.

**Large-scale simulations.** Within each digital cell, the chemotaxis network was simulated using stochastic methods (10, 30) with rates as in Table S1. Digital cells for each population were first adapted to a constant external concentration of aspartate of 1μM for more than 1000 s. Then, at time point t=0, cells were placed at z=0 in a linear gradient of attractant. This gradient ($10^{-8}$ M/μm) increases along the z direction ($L = 1\,\mu M$ at z=0) in a 3D infinite medium. The trajectories of individual cells were subject to rotational diffusion with diffusion coefficient $D_r = 0.062\,\text{rad}^2/\text{s}$ (33).

**Supporting Information.**

http://cluzel.uchicago.edu/data/emonet/arxiv_070531_supp.pdf

**Acknowledgments:** We thank H. Park and T. Shimizu for insightful discussions, C. Macal and M. North at Argonne National Laboratory for providing computing time on the Jazz Linux cluster and Eric Lyons for providing the stochastic simulator Stochastirator. This work was supported by NIH and by joint research funding under H.28 of the U. S. DOE Contract W-31-109-ENG-38.